\documentstyle[rep12]{article}
 \newlength{\dinwidth}
   \newlength{\dinmargin}
\def\reff#1{(\ref{#1})}
\def\Fzero{F_0^{\varepsilon\dagger}}
\def\Fdagger{F_\infty^{\dagger}}

\begin{document}
\newtoks\@stequation

     \def\subequations{\stepcounter{equation}
       \edef\@savedequation{\the\c@equation}
       \@stequation=\expandafter{\theequation}   
       \edef\@savedtheequation{\the\@stequation} 
       \edef\oldtheequation{\theequation}
       \setcounter{equation}{0}
       \def\theequation{\oldtheequation\alph{equation}}}

     \def\endsubequations{
       \setcounter{equation}{\@savedequation}
       \@stequation=\expandafter{\@savedtheequation}
       \edef\theequation{\the\@stequation}}

\thispagestyle{empty}
\setcounter{page}{0}
\begin{flushright}
\begin{tabular}{ll}
DFTUZ & 92.9 \\
FTUAM & 92-13\\
\end{tabular}
\end{flushright}
\vskip 1em
\def\foota{Departamento de F\'{\i}sica Te\'orica. Facultad de Ciencias.
           Universidad de Zaragoza. 50009 Zaragoza. Spain.
           e-mail: asorey@cc.unizar.es}
\def\footd{Departamento de F\'{\i}sica Te\'orica. Facultad de Ciencias.
           Universidad de Zaragoza. 50009 Zaragoza. Spain.
           e-mail: esteve@cc.unizar.es}
\def\footb{Institut de Physique Th\'eorique.
           \'Ecole Polytechnique F\'ed\'erale de Lausanne.
           CH-1015 Lausanne. Switzerland.
           e-mail: fernandez@eldpa.epfl.ch}
\def\footc{Departamento de F\'{\i}sica Te\'orica C-XI.
           Universidad Aut\'onoma de Madrid.
           Cantoblanco 28049 Madrid. Spain.
           e-mail: duncan@vm1.sdi.uam.es }
\begin{center}
{\LARGE {Rigorous Analysis of Renormalization Group Pathologies
         in the 4-State Clock Model}}
\vskip 3.5em
{\bf{\bf M. Asorey       }}\footnote{\foota}{\large{\bf,}}
{\bf {\bf J. G. Esteve   }}\footnote{\footd}{\large{\bf,}}
{\bf {\bf R. Fern\'andez }}\footnote{\footb}
{\bf {\bf and J. Salas   }}\footnote{\footc}
\vskip 1em
\vskip 4em
{\small {\bf Abstract}}
\end{center}
\vskip 0.5em
{\rm
We perform an exact renormalization-group analysis of
one-dimensional 4-state clock models with  complex interactions.
Our aim is to provide a simple explicit illustration of the behavior
of the renormalization-group flow in a system exhibiting a rich
phase diagram.  In particular we study the flow in the vicinity of
phase transitions with a first-order character, a matter that has been
controversial for years.  We observe that the flow is continuous and
single-valued, even on the phase transition surface, provided that the
renormalized Hamiltonian exist. The characteristics of such a flow
are in agreement with the Nienhuis-Nauenberg standard scenario,
and in disagreement with the ``discontinuity scenario''
proposed by some authors and recently disproved
by van Enter, Fern\'andez and Sokal for a large class of
models (with real interactions).
However, there are some points in the space of interactions for
which a renormalized Hamiltonian cannot be defined.
This pathological behavior is similar, and in some sense
complementary, to
the one pointed out by Griffiths, Pearce and Israel for Ising models.
We explicitly see
that if the transformation is truncated so as to preserve a Hamiltonian
description, the resulting flow becomes discontinuous and multivalued
at some of these points.
This suggests a possible explanation for the numerical results that
motivated the ``discontinuity scenario''.

\newpage

\baselineskip=18pt

\section{Introduction}

One of the major features of the Renormalization Group (RG)
theory is that it makes possible the description of the singular
critical behavior associated to second-order phase transitions in
terms of {\it smooth} RG transformations  on a
suitable space of Hamiltonians. By suitable we mean {\it local} (in
some mathematical sense that must be specified) and
translational-invariant Hamiltonians. The
thermodynamic singularities associated with critical phenomena
arise after an infinite number of RG steps, in the vicinity of a
RG fixed point \cite{wilson,wilson-kogut}.

The smoothness of the RG flow is also expected to  hold for other
values of the coupling constants. In particular, it was
conjectured by Nienhuis and
Nauenberg \cite{nienhuis} that it can also be smooth around points
where
first-order phase transitions occur. In their picture ({\it standard
scenario}) these points are in the domain of attraction of certain
fixed points---called {\it discontinuity fixed points}---that govern
the behavior of the system on the coexistence manifold.
These fixed points
have  relevant directions whose critical exponents are equal
to the dimensionality of the model $y = D$. The
singularities associated to first-order phase transitions
are obtained by an infinite iteration of
RG transformations around the discontinuity fixed points
\cite{nienhuis,klein,fisher},
in complete analogy with the case of second-order phase transitions.

On the other hand different authors
\cite{blote,lang,tony_z2,hasenfratz,decker} have exhibited
numerical evidence and arguments indicating that,
in disagreement with the picture advocated by the standard scenario,
the RG transformation could be discontinuous at the transition points
and could associate different renormalized Hamiltonians to
the different phases coexisting at the transition (multivaluedness).

Recently, van Enter-Fern\'andez-Sokal \cite{sokal,unpub} have
rigorously shown that
the second picture cannot hold for a large class of systems.
For classical variables  taking  values on a compact
manifold with real-valued local interactions (e.g. absolutely
summable Hamiltonians), they prove that---whenever defined---a
renormalization-group map associated to a local (real space)
RG prescription is single-valued and continuous on a suitable space
of Hamiltonians.
However, they have also shown that at---or in the vicinity of---a
transition
surface, a renormalized {\it local} Hamiltonian may not exist at all.
This pathology is called {\it non-Gibbsianness}\/, and was previously
pointed out by Israel \cite{israel} and, in some sense, previously
by Griffiths and Pearce \cite{griffiths}.
Its occurrence has been rigorously verified \cite{sokal,unpub} for the
Ising model in dimensions $D \geq 2$ at sufficient low temperature for
some
RG prescriptions (wich include decimation, some cases of majority
rule, block-averaging and Kadanoff transformations).

These results suggest that the discontinuities observed in
Refs.~\cite{blote,lang,tony_z2} are in fact an
artifact of the truncation of the renormalization scheme.
If the renormalized local Hamiltonian
exists, the size of the observed discontinuity should decrease
for smaller truncations.
On the other hand, it is plausible that the numerical manifestation
of non-Gibbsianness
be phase-dependent yielding an apparent discontinuity that
persists (possibly acquiring an oscillatory character) for
successively smaller truncations \cite{unpub}.
In relation to the validity of these explanations we mention the work
of Ref.~\cite{tony-salas}, where truncation errors were estimated for
Monte Carlo RG calculations of the two-dimensional Ising model
below the critical temperature.  They were found to be of the same
order of magnitude as the observed discontinuity.

The observations of the preceding paragraph do not apply to the
results
of Ref.~\cite{decker}, which are not obtained via a truncation scheme.
However, the method used there relies on the hypothesis that there is
only one
renormalized trajectory flowing away from the discontinuity fixed
point \cite{wilson2}. The reported discontinuity of the
renormalization
flow could hence be a consequence of the fact that this hypothesis
is not valid for the
two-dimensional Ising model which has two relevant operators:
one associated with the temperature ($= 1/\beta$) and
the other with an external magnetic field \cite{klein,tony-salas}.

In this paper we intend to clarify further the above issues
by analyzing very simple one-dimensional models that on the one
hand exhibit a rich phase diagram and on the other hand admit a RG
scheme that can be exactly computed.  The models belong to
a family of one-dimensional $q$-state clock
models with complex-valued Hamiltonians previously introduced by
some of us \cite{asorey}.  These models were intended as
simplifications
of quantum spin models of present interest in which either the
complex interactions are present {\it ab initio}\/ or they appear in
the effective Hamiltonian upon integration of fermionic degrees of
freedom.
Complex couplings arise, for instance, in the classical spin model
associated to a quantum Heisenberg
chain \cite{affleck,haldane,affleck91}, the effective model for the
quantum Hall
effect considered in \cite{pruisken,levine}, and the chiral Potts
model
which has been a recent focus of interest
\cite{chiralpotts1,chiralpotts2}
and is not unrelated to the model we analyze.

On the other hand, the presence of  complex interactions introduces
very interesting modifications in
the beautiful picture developed for traditional (real-interaction)
statistical mechanics---pure phases, ergodic decomposition of states,
independence of the boundary conditions for the thermodynamic potentials
\cite{ruelle,israelconvexity}.
%
%
This could merit further analysis, given
the role played by complex interactions
\cite{Lee-Yang,Griffiths,Fisher,Cardy,Gallavotti-Lebowitz} in the
study of several phenomena addressed by equilibrium statistical
mechanics: Lee-Yang
singularity \cite{Lee-Yang,Griffiths,Fisher,Cardy},
%
%
metastability
effects \cite{isakov,milos}, high-$T$ analyticity
\cite{galsol,isr,dobmar} and
low-$T$ smoothness \cite{isakov} of the free energy, and deformations
of the phase diagram at low temperature \cite{milos}.
%
%
An important change incorporated by complex interactions is, of course,
the enrichment of the phase diagram:  the nearest-neighbor
one-dimensional models studied here do present phase transitions, a
feature totally absent
in real-interaction models of comparable simplicity \cite{teoremas}.

The models with complex interactions that we consider here have several
desirable properties \cite{asorey}:  They possess a Hermitean transfer
matrix which is positive-definite
 for a wide range
of coupling constants that includes phase transition points.
This last property
makes these models useful to generate unitary quantum-mechanical
systems in the continuum limit \cite{asorey-esteve,borgs}.
%
%
Although the models have complex interactions the free and internal energies
are real for periodic boundary conditions.
In addition, these models present a nontrivial phase diagram
characterized
by a manifold of points where the leading eigenvalues of the transfer
matrix cross.
At these transition points, the energy density is in general
discontinuous, so the transition can be catalogued as first-order;
but at the same time for many values of the couplings constants the
correlation length diverges with a critical exponent $\nu = 1$.
Therefore, the surface formed by these points is also a critical
surface. On the other hand, inside the transition manifolds there run
curves where the thermodynamic behavior is, in some sense, even more
singular. They correspond to the points of the phase diagram where the
partition function is zero for (a sequence of)
arbitrarily large volumes, and suitable boundary conditions.
For lack of better or established nomenclature, we shall call these
points---which have no counterpart in the phase diagrams of real
interactions---``Lee-Yang-type'' (LYT) points.
This type of singularities is familiar to people studying metastability
\cite{isakov,milos}.
In principle, they correspond to
singularities for the {\it finite}-volume free energies and its
existence is not enough to
infere the non-analyticity of the {\it infinite}-volume free
energy\footnote{Simple example:  The singular functions
$f_n(x) = \ln\left(1- {ix\over n}\right)$ converge to the perfectly
analytical zero function.  We thank Alan Sokal for clarifications
regarding this point.}---it
only rules out the usual analyticity proof pionered by Lee and
Yang \cite{Lee-Yang}.
For real interactions, the question of whether there is
a singularity in the infinite-volume limit is related to the
possibility of analytically continuing the (infinite-volume)
free energy in the presence of metastable states \cite{isakov}.
In our model, however, the LYT singularities do have infinite-volume
consequences:  they belong to transition surfaces and in addition,
as we shall see, the free energy presents some further singular
behavior at these points.

We believe that all these attributes make the clock models a useful
laboratory  for the study of RG flows.  Since the models
are one-dimensional  and have  nearest-neighbor interactions the
natural candidate for the RG prescription is decimation.  Such
transformations have been used for many one-dimensional Ising
models with real local couplings, see e.g. \cite{Fisher-Nelson}.
However, these were models  without first-order transition points. The
first application to a model with this type of transitions has been
presented recently \cite{clock3}.
The flow of the decimation transformation
is comparatively simple for these models:  it involves a finite
number of parameters, so it can be computed exactly without any
truncation approximation.
We present here the results for the 4-state clock model
subjected to decimations with blocks of size even.
%
%
The calculations reveal a number of instructive features which, we
think, bear some light on present controversies on the properties of
RG transformations.  Let us summarize our main observations:
\smallskip

1)  Our results are in agreement with
the standard scenario of Nienhuis and Nauenberg and the
van Enter-Fern\'andez-Sokal theorems (even when, rigorously speaking,
our model does not satisfy the hypothesis of these theorems because
it has complex interactions):  {\it Whenever defined}\/,
the flow is continuous and the behavior of the model on the surface of
first-order transitions is determined by (three) discontinuity fixed
points
lying on this surface.  One of these fixed points has two relevant
directions characterized by the same critical exponent $y=D=1$, as
predicted by the standard scenario.
The other two discontinuity fixed points are non-Gibbsian, and hence
the flow is singular.

2)  In some regions of the phase diagram we observe pathologies
similar to those pointed out
by Griffiths, Pearce and Israel \cite{israel,griffiths}:  Already for
the first renormalization step the
renormalized Hamiltonian fails to exist.  The reason for this
non-Gibbsianness
is, however, different from the one observed for real interactions.
While for the latter the renormalized measure fails to be
(quasi)local \cite{sokal,unpub,israel}---roughly speaking it looks as
if the
coupling constants proliferate in an uncontrollable manner---in the
present
case the renormalized measure gives zero weight to (open) sets of
configurations.  This corresponds to one of the coupling constants
attaining the value $+\infty$.
In a more abstract language, the renormalized measure fails to
satisfy uniform non-nullness.  Such a pathology cannot happen in the
renormalization
of real interactions, except if the renormalization prescription
itself
excludes some configurations.  Moreover, for real interactions
quasilocality
and uniform nonnullness are necessary and sufficient conditions for
a measure to be Gibbsian \cite{kozlov} (in the complex case the
matter
is much more involved).  In this sense, we can say that the
pathologies
observed in the present example are complementary to the ones
discussed in Refs.~\cite{sokal,unpub,israel,griffiths}.
We observe that if at some of these points the renormalized
Hamiltonian is
``truncated'' by ignoring the coupling constant that acquires an
infinite
value, the resulting transformation becomes discontinuous and
multivalued.
In this sense we could say that in this example the
non-Gibbsianness
produces a ``discontinuity scenario'' as a result of truncation.

3)  We can distinguish two different ``degrees'' of pathological
behavior.
In most of the pathological points, the non-Gibbsianness brought by
the RG transformation is ``recoverable'': a further iteration of the
RG transformation restores the Gibbsianness, and this Gibbsianness is
preserved under additional iterations.  Equivalently, the points are
pathological for one particular even-block decimation scheme but not
pathological for all the others.  In constrast, at the points
LYT---and only there---the non-Gibbsianness is
not restored by a further renormalization.  The points LYT are,
therefore, pathological for {\em all} even-block decimation schemes.
We observe that the LYT singularities correspond to points where
all the
eigenvalues of the transfer matrix are doubly degenerated (in
absolute
value).  While we do not fully understand the meaning of this
observation,
we remark that a similar degeneracy seems to be present in the
critical curve of the two-dimensional Ising model \cite{minlossinai}.

4)  The RG flow exhibits the following characteristics:  It
involves two families of models, both of them formed by 4-state clock
models parametrized by three real numbers denoted $J$, $J_1$ and
$\varepsilon$, but differing in the value of a discrete parameter $m$
that gives an imaginary part to the couplings.
One family is a (transfer-matrix-Hermiteanness-preserving) complex
extension of the other.  For each decimation prescription,
each family can be divided in two sections:  an open region of the
parameter space that we call the black-hole section, and its
complement, the
non-black-hole section.  The RG transformation maps models in the
{\em non-black-hole} section into models of the {\em same} family,
and models in the {\em black-hole} section into models of {\em the
other}
family.  The points where the RG transformation has a ``recoverable''
pathology are precisely
those of the boundary between black-hole and non-black-hole sections
that are not LYT singularities.
These boundaries change with the
decimation scheme; the only points common to all of them are curves
of LYT singularities.

5)  Each family of models exhibits seven different fixed points (plus
periodic repetitions), but three of them attract the majority of the
models with finite couplings:  a ``high-temperature'' (zero
correlation
length) fixed point, and the two non-Gibbsian discontinuity fixed
points
located on the critical surface.  Models outside the critical surface
are attracted by the high-temperature fixed point---of the same or the
other family, depending on which side of the critical surface the
initial
models are.  On the other hand, models on the critical surface are
attracted by the non-Gibbsian critical fixed points of the same or
the other family, depending on wether initially the models are inside
or outside the black-hole region.  There is only a curve of points
inside
the critical surface that are attracted to the Nienhuis-Nauenberg
fixed point
mentioned above.  Furthermore, we can explain the ``terminal''
pathologies affecting the points LYT in terms of strongly attracting
invariant planes of points:  By a {\em single} renormalization
transformation
the points LYT are mapped either into a non-Gibbsian critical fixed
point or into an invariant plane of non-Gibbsian models.
\smallskip

The paper is organized as follows. In Section 2 we define the 4-state
clock model and study its phase structure in detail. The RG analysis
is mainly explained in Section 3. We see that in some regions of the
interaction space an analytic continuation of the RG equations is
needed (passing to another Riemann sheet).  This is performed in
Section 4 and consists in adding an
additional parameter to the Hamiltonian. The pathologies observed are
discussed in Section 5.  In Section 6 we present some final comments.

\section{Phase Structure of the 4-State Clock Model}

A one-dimensional $q$-state clock model is defined by
a classical variable $\vec{s}_n$ (``spin'') fluctuating among
the $q$ roots of the unity
\begin{equation}
\vec{s}_n = \left( \cos \left( \frac{2\pi p_n}{q} \right) ,
                   \sin \left( \frac{2\pi p_n}{q} \right)  \right) ; \
                   p_n=0,\ldots,q-1
\end{equation}
on the sites  $n$ of a one-dimensional chain.
%
%
We consider the Hamiltonian
\begin{equation}
   -\beta {\cal H} =
                        \sum_{n=1,r=1}^{N,[\frac{q}{2}]}
 \left\{  J_{r-1} \cos \left( \frac{2 r \pi}{q} (p_n-p_{n+1}) \right)
 + i\epsilon_r \sin \left( \frac{2 r \pi}{q} (p_n-p_{n+1}) \right)
 \right\},
\end{equation}
with $J_r$ and $\varepsilon_r$ real parameters, and
where $[q/2]$ denotes the integer part of $q/2$. This type of
Hamiltonian is a natural generalization of the Hamiltonians studied
in \cite{asorey},
with additional interactions introduced so to have a system of
couplings closed under decimation transformations, that is,
%
%
with as many independent couplings (including the
free energy normalization) as RG equations \cite{clock3}.
In the case $q = 4$ the Hamiltonian is
\begin{eqnarray}
-\beta {\cal H}
                &=&   \sum_{n=1}^N  \left\{
    J\cos \left( \frac{\pi}{2}(p_n-p_{n+1}) \right) +
  J_1\cos      (       \pi    (p_n - p_{n+1})) \right. \nonumber \\
              &  & \qquad+   \left.
i\varepsilon \sin \left( \frac{\pi}{2}(p_n-p_{n+1}) \right)
\right\}.
\label{clock}
\end{eqnarray}
(We have written ${ J,\varepsilon}$ instead of ${ J_0,\varepsilon_1}$
to conform with previous works
\cite{asorey,clock3}.)
%
%
The transfer matrix of this model reads
\begin{equation}
T  = \left(
\begin{array}{rrrr}
e^{J+J_1}               &   e^{-i\varepsilon - J_1}    &
e^{-J+J_1}              &   e^{i\varepsilon - J_1}     \\
e^{i\varepsilon - J_1}  &   e^{J+J_1}                  &
e^{-i\varepsilon - J_1} &   e^{-J+J_1}                 \\
e^{-J+J_1}              &   e^{i\varepsilon - J_1}     &
e^{J+J_1}               &   e^{-i\varepsilon - J_1}    \\
e^{-i\varepsilon - J_1} &   e^{-J+J_1}                 &
e^{i\varepsilon - J_1}  &   e^{J+J_1}
\label{tmatrix}
\end{array}
\right)
\end{equation}
and its eigenvalues are
\begin{equation}
\begin{array}{l}
  \lambda_0 =
        2e^{J_1} \cosh J + 2 e^{-J_1}\cos \varepsilon \\
  \lambda_2 =
        2e^{J_1} \cosh J - 2 e^{-J_1}\cos \varepsilon \\
  \lambda_1
        = 2e^{J_1} \sinh J + 2 e^{-J_1}\sin \varepsilon \\
  \lambda_3
        = 2e^{J_1} \sinh J - 2 e^{-J_1}\sin \varepsilon \;.
\end{array}
\label{eigenvalues}
\end{equation}
The eigenvectors of the transfer matrix are spin waves: the
eigenvector for the eigenvalue $\lambda_k$ has components
\begin{equation}
\omega_m^{(k)} =
e^{i \pi k m /2}\qquad (m = 0, \ldots, 3).
\label{eigenvectors}
\end{equation}
{}From \reff{eigenvalues}--\reff{eigenvectors} it is easy to obtain
explicit expresions for arbitrary powers of the transfer matrix:
\begin{equation}
(T^n)_{q\,q'} = (1/4) \sum_{k=0}^3 (\lambda_k)^n \,\exp[i\pi
k(q-q')/2]\;.
\label{tm}
\end{equation}
%
%
%
%
With this formula we can calculate all the statistical mechanical and
thermodynamic properties of the model.

Before discussing the phase diagram we observe the following
symmetries of the eigenvalues \reff{eigenvalues}:
\begin{eqnarray}
\lambda_{0,2}(\pi/2-\varepsilon) =\lambda_{2,0}(\varepsilon+\pi/2),
  &  &
\lambda_{1,3}(\pi/2-\varepsilon) =\lambda_{1,3}(\varepsilon+\pi/2)
\label{pi2}\\
\lambda_{0,2}(-\varepsilon) =\lambda_{0,2}(\varepsilon),
  &  &
\lambda_{1,3}(-\varepsilon) =  \lambda_{3,1}(\varepsilon)
\label{me}\\
\lambda_{0,2}(-J) = \lambda_{0,2}(J),
  &  &
\lambda_{1,3}(-J) = -\lambda_{3,1}(J)
\label{mj}
\end{eqnarray}
Due to them, we only need and will describe the phase diagram in the
region $0\leq \varepsilon\leq \pi/2$, $J\geq 0$.  The diagram on
$\pi/2\leq\varepsilon\leq \pi$ can be obtained by
reflections on the plane $\varepsilon=\pi/2$ together with the
interchange $\lambda_0 \longleftrightarrow \lambda_2$  (symmetry
\reff{pi2}); the diagram on $-\pi\leq\varepsilon\leq 0$ by reflections
on the plane $\varepsilon=0$ plus the interchange
$\lambda_1 \longleftrightarrow \lambda_3$ (symmetry \reff{me}); and
the diagram for other values of $\varepsilon$ follows from the
$2\pi$-periodicity.  The diagram for $J\leq 0$ is obtained by
reflections on the plane $J=0$ followed by the interchange
$\lambda_1 \longleftrightarrow -\lambda_3$ (symmetry \reff{mj}).
As an example,
Fig.~\ref{fig1} shows the dependence of
those eigenvalues on the coupling $\varepsilon$ of the imaginary term
for some choice of $J$ and $J_1$.

In the thermodynamic limit ($N \rightarrow \infty$) the behavior of the
system is regulated by the leading eigenvalues.  By ``leading'' we mean
largest {\em in absolute value}\/.
For those values of the parameters for which there is only one leading
eigenvalue, say $\lambda_{k_0}$, we can see from \reff{tm}
that the free energy density for even chains takes the value\footnote{This
also holds for odd chains with an appropiate branch
choice in the logarithm definition.}  $\log|\lambda_{k_0}|$
and all the expectations are independent of the boundary conditions.
By all accounts, these correspond to regions with only one phase
present. Crossing points of two leading eigenvalues
correspond to transition points, (see Fig.~\ref{fig1}).

In the region $0\leq \varepsilon\leq \pi/2$, $J\geq 0$, the
transition surface is composed of three pieces (Fig.~\ref{fig2}):
\begin{itemize}

\item The surface $C_{0,1}$ defined by the condition
$\lambda_0=\lambda_1> \max{(\lambda_2,|\lambda_3|)}$:
\begin{equation}
C_{0,1} \;=\; \left\{ (J,J_1,\varepsilon) :
0\leq \varepsilon\leq \pi/2\;,\; J\geq 0 \;{\rm and}\;
  \varepsilon = \frac{\pi}{4} +
               {\rm arcsin} \left( \frac{e^{-J+2J_1}}{\sqrt{2}}
                            \right) \right\}.
\label{c01}
\end{equation}

\item The surface $C_{1,-3}$ defined by the condition
$\lambda_1=-\lambda_3> \max{(\lambda_0,\lambda_2)}$:
\begin{equation}
C_{1,-3} \;=\; \left\{ (J,J_1,\varepsilon) :
\pi/4 \leq \varepsilon\leq \pi/2\;,\; J=0 \;{\rm and}\;
  \varepsilon \geq \frac{\pi}{4} +
             {\rm arcsin} \left( \frac{e^{2J_1}}{\sqrt{2}}
                         \right) \right\}.
\label{c1-3}
\end{equation}

\item The surface $C_{0,2}$ defined by the condition
$\lambda_0=\lambda_2> \max{(|\lambda_1|,|\lambda_3|)}$:
\begin{equation}
C_{0,2} \;=\; \bigl\{ (J,J_1,\varepsilon) :
\varepsilon= \pi/2 \;{\rm and}\; 0\leq J\leq 2 J_1 \bigr\}.
\label{c02}
\end{equation}
\end{itemize}
In addition, the degenerated planes $J=\infty$  and
$J_1=\infty$  can also be
considered transition surfaces, but we are more interested in the
properties of models with finite couplings.

%
%

It is trivial to see that the energy density is
discontinuous at the transition points, which implies that the
system undergoes a first-order phase transition.
However, the correlation length also diverges at the points of the
surface $ C_{0,1}$  and this
transition surface can also be considered as a critical surface.
This can be seen, for instance, by computing the correlation of two
spin variables
%
%
\begin{equation}
   \left\langle \vec{s}_n \cdot \vec{s}_{n+m} \right\rangle =
       \frac{1}{2}  \left[
       \left( \frac{\lambda_{k_0-1}}{\lambda_{k_0}} \right)^m +
       \left( \frac{\lambda_{k_0+1}}{\lambda_{k_0}} \right)^m
                                       \right],
\label{correlation}
\end{equation}
where $\lambda_{k_0}$ is the leading eigenvalue.
Eq.~\reff{correlation} implies
that the correlation length diverges at the transition points $
C_{0,1}$ as $\xi(\beta)=(\beta^{-1}-\beta^{-1}_c)^{-1}$ and hence
one obtains critical exponents $\nu = \eta = 1$.
The correlation
length also diverges when $J$  goes to infinity and, in fact,
in the limit $J \rightarrow \infty$ the discontinuity of the energy
density goes to zero and the transition becomes purely second order.

An important feature brought by the use of complex couplings is the
presence of ``Lee-Yang-type'' (LYT) singularities.
We so denominate those values of the
couplings for which there exists a sequence of volumes and boundary
conditions giving a partition function equal to zero. For the present
models this condition requires the existence of
a sequence of powers of the transfer matrix
with some entry (which could be different for different powers)
equal to
zero.  Therefore, the locus of the points LYT can easily be
obtained from \reff{tm}
together with the form \reff{eigenvalues} of the eigenvalues
$\lambda_k$.
The conclusion is that, for finite couplings, there are only two
curves of points LYT (again, for the region
$0\leq\varepsilon\leq\pi/2$, $J\geq 0$; for the rest of the phase
diagram one must proceed as commented below \reff{mj}):
\begin{itemize}
\item The line defined by the equations $\lambda_0=\lambda_2$ and
$\lambda_1=-\lambda_3$; that is:
\begin{equation}
{\rm LYT1}= \left\{ (J,J_1,\varepsilon) : J=0,
\varepsilon=\pi/2\right\}. \label{lyt1}
\end{equation}
For couplings on this line,
\begin{equation}
(T^n)_{q\,q'} = 0 \quad \iff \quad n \hbox{ is even and }
|q-q'|=1,3\;. \label{tlyt1}
\end{equation}

\item The curve defined by the equations $\lambda_0=\lambda_1$ and
$\lambda_2=-\lambda_3$; that is:
\begin{equation}
{\rm LYT2}= \left\{ (J,J_1,\varepsilon) \in C_{0,1} :
           \cosh 2J \,\cos 2 \varepsilon = -1 \right\}
\label{lyt2}
\end{equation}
On this curve,
\begin{equation}
(T^n)_{q\,q'} = 0 \quad \iff \quad n \hbox{ is even and } |q-q'|=2
\label{tlyt2}
\end{equation}

\end{itemize}
(See Fig.~\ref{fig2}).
We remark that these are the only solutions if we search for
a {\em sequence} of volumes with zero partition functions.
In addition there are whole surfaces in the coupling-constant space
formed by points for which the
partition function is zero for {\em some} volume and boundary
condition.
These points are precisely those of the boundary $\partial B_{2k}$ of
the black-hole region (see next section) for some even-block
decimation.
We note, in passing, that the diagonal entries $(T^{2n})_{qq}$ are
always nonvanishing, in fact strictly positive (for finite
couplings).

Both curves LYT are contained in transition surfaces:  LYT2 is in
$C_{0,1}$
and LYT1 is in $C_{0,2}$ for $J_1\geq0$ and in $C_{1,-3}$ for
$J_1\leq0$.
In fact, the curves correspond to those points of the transition
manifold where {\em all} the eigenvalues
become pairwise degenerate (in absolute value).
The curves intersect at the point
\begin{equation}
P_{\rm LYT} = (J=0\,, J_1=0\,, \varepsilon=\pi/2)
\label{plyt}
\end{equation}
which correspond to all the eigenvalues having the same absolute value
(maximun degeneration).  This
point is extremely singular for the flow of the decimation
transformations (Section \ref{pato}).
\medskip

Within the critical manifold $ C_{0,1} $ we observe some thermodynamic
features
very different from those of real interactions:  the infinite-volume
free energy acquires a dependence on the boundary conditions, and
no boundary condition produces truncated correlations with decaying
behaviour. Let us denote $f_{q,q'}$ the free
energy for boundary conditions $q$ on the left and $q'$ on the right.
{}From \reff{tm} we obtain, for example, that
in the surface $C_{0,1}$,  $f_{q,q'}=\log\lambda_0$ (=$\log\lambda_1$)
for
all boundary conditions $q,q'$ {\em except} if $q'=q\pm2$.  In that
case the free energy takes the values $\log\lambda_2$
or $\log|\lambda_{3}|$ depending on which is the subleading
eigenvalue. The transition between
these two values takes place at the curve LYT2---both non-leading
eigenvalues become equal in absolute value---where $f_{q,q\pm2}$ can
only be defined by considering volumes of odd size.
\medskip

For the construction of quantum-mechanical systems via scaling limit,
only the region of parameters for which the model
satisfies the reflection-positivity condition is of interest.
For odd chains, this condition is satisfied for all values of the
parameters because the transfer matrix is
Hermitean \cite{asorey}. For even chains the reflection-positivity
holds only if all the eigenvalues of the transfer matrix are
non-negative, that is, for $J\geq J_{RP}(J_1,\varepsilon)$, with
\begin{equation}
   J_{RP}(J_1,\varepsilon) =
   \max \biggl\{ {\rm arccosh} (e^{-2J_1} |\cos \varepsilon|),
        {\rm arcsinh} (e^{-2J_1} |\sin \varepsilon|) \biggr\}
\label{rp}
\end{equation}
(Figs.~\ref{fig3} and \ref{fig4}).
%
%
In this work we do not restrict ourselves to the
reflection-positive region; if we did we would miss the curves LYT
which are the most interesting regions from the point of view of the
pathologies of the decimation transformation.

\section{Renormalization Group Flow}

A single one-spin decimation transformation in the system described
above yields the following Renormalization-Group equations
\begin{eqnarray}
J' &= & \frac{1}{2} \log  \left\{
        \frac{1+e^{4J_1}\cosh 2J }
             {e^{4J_1}+ \cos 2\varepsilon}  \right\}  \nonumber \\
J_1'&=& \frac{1}{4} \log  \left\{
        \frac{ (1+e^{-4J_1}\cos 2\varepsilon)
               (1+e^{4J_1}\cosh 2J) }
             {2 (\cosh 2J + \cos2\varepsilon)}  \right\}
\label{decimation} \\
\varepsilon'&=& \frac{1}{2} \arccos \left\{
        \frac{ 1+\cosh 2J \cos 2\varepsilon }
             {\cosh 2J+\cos2 \varepsilon}       \right\}  \nonumber
\end{eqnarray}
where the primes denote the renormalized quantities.
To decide the quadrant of $\varepsilon'$, the last equation must be
complemented with
\begin{equation}
\tan\varepsilon' \;= \; \tan\varepsilon\,\tanh J\;.
\label{quadrant}
\end{equation}
This is the complete exact RG flow except that, as usual, we have
omitted the renormalization of the coupling associated with the
identity operator. This corresponds to a spin-independent constant
added
to the Hamiltonian, which is therefore irrelevant for the analysis of
correlation functions and thermodynamic potentials.
%
%
By iteration of \reff{decimation} we can understand the flow of all
decimation transformations of even blocks (decimation of an odd number
of spins $=$ even powers of $T$).

Starting with the coarser features of the flow, we first observe
that
all such transformations map the region $J < J_{RP}(J_1,\varepsilon)$
(non-reflection-positive region) onto the
complementary region $J > J_{RP}(J_1,\varepsilon)$ by a single RG
transformation \cite{clock3}. This is due to the fact that the matrix
$T^2$ is always reflection positive.  In particular all the fixed
points lie in the reflection-positive region.
We then notice that the flow defined by
\reff{decimation}--\reff{quadrant}
exhibits a periodicity of $\pi$ in $\varepsilon$ and very distinct
symmetries with respect to the change of sign of $J$ and
$\varepsilon$.
We conclude that it is enough to study the flow in the region of
Fig.~\ref{fig2} --- ($J\geq 0$ and $0\leq \varepsilon \leq \pi/2$) ---
and so we will in the sequel.  The flow for $J\geq 0$ and
$-\pi/2\leq\varepsilon\leq 0$ is an identical copy, while the points
with
$J\leq 0$ change quadrant:  their renormalized couplings are the
same as
those for $|J|$, but with opposite sign for $\varepsilon'$.  Having
said
this, we can forget about the quadrant-fixing equation
\reff{quadrant}.

Another conspicuous feature of the RG transformation
\reff{decimation} is the presence of a large region
in the coupling constant space where the renormalized couplings are not
real numbers.  We shall call this region the black
hole $B_2$ (see Figs.~\ref{fig3} and \ref{fig4}):
\begin{equation}
B_2= \left\{ (J,J_1,\varepsilon):e^{4J_1}<- \cos 2\varepsilon,\
\varepsilon > \frac{\pi}{4} \right\}
\label{black}
\end{equation}
Notice that the
intersection of the black hole region with the transition surface is
non-empty, even in the region where reflection positivity holds
(Fig.~\ref{fig3}).
While \reff{black} corresponds to the decimation of alternated spins,
decimations of chains of $k$ spins also exhibit a corresponding
black-hole region $B_{2k}$.  In Section \ref{ablack} we discuss how this
region changes with the order $k$ of the decimation.  In the black-hole
region we must
%
%
extend analytically \reff{decimation} into the complex space
$(J,J_1,\varepsilon) \in \mbox{{\bf C}}^3$.
However, this extension must be performed so as to preserve a
physically highly desirable property that motivated the initial
restriction to real $(J,J_1,\varepsilon)$:  a real free energy.
For this we extend \reff{decimation} in such a way that the transfer
matrix $T$ remains Hermitean. This extension will be done in detail in
Section \ref{ablack}.

%

The boundary
\begin{equation}
\partial B_2= \left\{ (J,J_1,\varepsilon):e^{4J_1} = - \cos
2\varepsilon,\ \varepsilon > \frac{\pi}{4} \right\}
\end{equation}
of the black hole is a pathological region for the RG flow:
at these points the renormalized couplings take
the value $J'=+\infty$, $J'_1=-\infty$.  As we discuss in more detail
in Section \ref{pato},
this implies that the renormalized measure gives zero weight to some
sets of configurations, a property incompatible with a Boltzman
probability weight (exponentials are never zero).  The renormalized
Hamiltonian simply does not exist.
There is another pathological region, namely the line LYT1 ($J=0,
\varepsilon=\pi/2$), where $J'_1=+\infty$ and $\varepsilon'$ is
undefined.  In Section \ref{pato} we shall discuss more extensively
these pathologies.

In the rest of the coupling constant space (including the points
on the transition surface not in $\partial B_2$ or LYT1), the flow
is well
defined:  to every point the transformation \reff{decimation}
associates a unique renormalized Hamiltonian.
Such a non-pathological flow will be described in the remaining
part of this section.
\medskip

Due to the special features of the decimation procedure we can
easily locate RG invariant surfaces, that is, surfaces that are
mapped onto themselves by a RG transformation.
One of them is the reflection positivity interface defined
by \reff{rp} which is defined by the condition that the smallest
eigenvalue of the transfer matrix be equal to zero.  This condition is
invariant under a decimation transformation, a fact that can be more
easily seen if we write this transformation in the form
\begin{equation}
   \lambda_k^2(J,J_1,\varepsilon) = \lambda_k(J',J_1',\varepsilon')
\,L( J',J_1',\varepsilon'),
\label{better}
\end{equation}
where $L$ is the (strictly positive) overall factor associated to the
renormalization of the free energy.  The same expression \reff{better}
shows that equalities and inequalities among the {\em absolute values}
of
eigenvalues of $T$ are also preserved by decimations of any even
block, and,
therefore, that the reflection-positive part of the transition
surfaces
%
%
are invariant under these transformations.
%
%
In particular, the surfaces $C_{0,1}$, $C_{0,2}$, $J=\infty$ and
$J_1=\infty$ are RG invariant.
The intersections of two
invariant surfaces are RG invariant curves, that is,
renormalized trajectories. Some of them are drawn in
Fig.~\ref{fig5}a.

These renormalized trajectories intersect at the RG fixed points
(see Figs.~\ref{fig4} and \ref{fig5}a). They are listed in Table
\ref{table1}.
{}From \reff{correlation} it can be seen that the correlation length
diverges at $F_c$, $F_c^\ast$, $\Fzero$ and $F_0^\varepsilon$.

\begin{table}[h]
\centering
\begin{tabular}{|lll|}  \hline
   &  &         \\
 Point           & Position ($J,J_1,\varepsilon$) &
 Type            \\
   &  &         \\
  \hline
   &  &         \\
$F_\infty$       & $(0,0,0)$                      &
 stable       \\
$\Fdagger$      & $(0,\infty,0)$                  &
saddle point     \\
$F_c$            & $(\frac{1}{2}\log 3,\frac{1}{4}\log 3,\pi/2)$ &
saddle point    \\
$F^\ast_c$       & $(\infty,-\infty,\pi/4)$       &
saddle point     \\
$F^\dagger_c$    & $(0,\infty,\pi/2)$             &
saddle point     \\
$\Fzero $            & $(\infty,\infty,\varepsilon)
        \;;\;\varepsilon \in [0,\pi/4])$  &
unstable        \\
$F^\varepsilon_0$& $(\infty,\frac{1}{4} \log\cos2\varepsilon
            ,\varepsilon) \;;\;\varepsilon \in [0,\pi/4])$   &
saddle point  \\
   &  &         \\
  \hline
\end{tabular}
\caption{Fixed points for even-block decimations.} \label{table1}
\end{table}


The linearization of the RG equations \reff{decimation} around the
different fixed points yields the matrices
\begin{equation}
\begin{array}{ll}
  &   \\
L_{F_\infty}  = \left( \begin{array}{rrr}
                       0  &  0  &  0 \\
                       0  &  0  &  0 \\
                       0  &  0  &  0  \end{array} \right)    &
L_{F_c}       = \left( \begin{array}{rrr}
                       \frac{2}{3}  & -\frac{4}{3}  &  0 \\
                      -\frac{4}{3}  &  \frac{2}{3}  &  0 \\
                        0  &  0  &  2  \end{array} \right)    \\
  &   \\
L_{\Fdagger} = \left( \begin{array}{rrr}
                       0  &  0  &  0 \\
                       0  &  1  &  0 \\
                       0  &  0  &  0  \end{array} \right)    &
L_{\Fzero }       = \left( \begin{array}{rrr}
                       1  &  0  &  0 \\
                       0  &  1  &  0 \\
                       0  &  0  &  1  \end{array} \right)    \\
  &   \\
L_{F_0^\varepsilon} = \left( \begin{array}{ccc}
                  1  &  1        & \phantom{-}
                                     \frac{1}{2} \tan2\varepsilon \\
                  1  &\frac{1}{2}&  -\frac{1}{4} \tan2\varepsilon \\
                  0  &  0  &  1   \end{array} \right),    &
                                                             \\
  &   \\
\end{array}
\label{linearization}
\end{equation}
whose eigenvalues ($\zeta_i$) and eigenvectors determine
the critical exponents ($y_i = \log \zeta_i / \log 2$) and
the main properties of the renormalized trajectories.

The fixed point $F_\infty$ is completely attractive in all directions.
The point $\Fdagger$ behaves as an attractor with respect to
perturbations
along the $J$-axis and the $\varepsilon$-axis, but has a marginal
direction ($y = 0$) along the $J_1$-axis. The latter is
actually a relevant direction, as can be checked numerically, and
flows toward the infinite-temperature fixed point $F_\infty$.

There are three fixed points on the transition manifold:  $F_c$,
$F_c^\dagger$ and $F_c^\ast$.  The first one
has features predicted by the standard scenario:  It is a saddle
point with one irrelevant and two relevant directions, both with
critical exponents equal to $y =  D = 1 $.
Of the relevant directions, one corresponds to perturbations along the
$\varepsilon$-axis with the flow escaping towards the fixed point
$F_\infty$, while the other is tangent to the renormalized trajectory
$J = J_{RP}(J_1,\pi/2)$ where the flow moves towards $F^\dagger_c$ and
$F_c^\ast$.  The flow towards $F_c^\ast$ is along the renormalized
trajectory defined by the intersection between the
reflection-positivity
interface and the transition surface $C_{0,1}$ (given by \reff{rp}
and \reff{c01} respectively); the flow towards
$F_c^\dagger$ is along the line $J = J_{RP}(J_1,\pi/2)$.
The irrelevant direction is along the RG trajectory
$J = 2J_1$, $\varepsilon = \pi/2$.

The other two fixed points on the transition surface
correspond to singularities of the flow (they are non-Gibbsian)
and therefore the RG transformations can not be linearized around
them. We have analyzed the flow numerically and we
conclude that $F_c^\dagger$ is attractive for those transition
points satisfying $J < 2J_1$ (triangular-looking cusp in the surface
of Fig.~\ref{fig5}a), while $F_c^\ast$
attracts the rest of points of the transition surface.
In both fixed points there is one relevant direction  that flows away
from  the transition surface, whose critical exponent is not well
defined.

Finally we have two (``zero-temperature'') fixed-point lines $\Fzero
$, and $F_0^\varepsilon$.
There are three marginal directions at each point of $\Fzero $, but
only the one tangent to the line $\Fzero $ itself is truly marginal.
The other two are actually relevant and the fixed points become
unstable.
Each of the points of $F_0^\varepsilon$ has one irrelevant
perturbation
($y = -1$) along the $(2,-1,0)$ direction and two marginal ones:
one, truly marginal, tangent to the line $F_0^\varepsilon$ itself,
and the other given by the lines
$J_1 = \infty$, $\varepsilon \in [0,\pi/4]$, which are RG
trajectories flowing from $\Fzero $ to $F_0^\varepsilon$.
At these fixed points one can take the continuum limit. The
existence
of these lines of fixed points leads to different quantum mechanical
systems in the scaling limit \cite{asorey-esteve}.
\medskip

We now analyze the flow outside of the black-hole section $B_2$, to
follow
the points inside the black hole we need an analytic extension to be
discussed
in Section \ref{ablack} (See Figs.~\ref{fig5}a,
\ref{fig6}a and \ref{fig6}b).
For reading the rest of this section we recommend to look at the
Figures
\ref{fig4}, \ref{fig5}a, \ref{fig6}a and \ref{fig6}b as many times as
needed. We start with
the flow inside the invariant planes.  There are four of them,
corresponding
to $\varepsilon=0$, $\varepsilon=\pi/2$, $J_1=\infty$ and $J=\infty$.
{}From \reff{decimation} we see that $F_\infty$ attracts all points in
the
invariant plane $\varepsilon = 0$, except those sitting at the line
$J_1 = \infty$ which flow to $\Fdagger$. The fixed points $\Fzero $
are completely unstable.
In the plane $\varepsilon = \pi/2$ the situation is less simple:
$F_c$ attracts the transition points on the renormalized
trajectory $J = 2J_1$ and $F_c^\dagger$ is attractive for all
transition
points above this line ($J < 2J_1$). The remaining points in this
plane flow toward the black hole region in a finite number of RG
steps, except those of the boundary $\partial B_2$ which go to the
plane $J=\infty$ (at its intersection with $J_1=-\infty$) in one
renormalization step.
In the plane $J_1 = \infty$ all points are attracted by $\Fdagger$,
except those belonging to the line $\varepsilon = \pi/2$ which flow
toward $F_c^\dagger$. We remark that the points lying on the
line $J = 0$ are mapped to $\Fdagger$ in a single RG step.
Finally, in the plane $J = \infty$ every line $\varepsilon =
{\rm constant}$ is a trajectory of the renormalization transformation.
When $\varepsilon \in [\pi/4,\pi/2]$ they flow towards the black hole
region, while the points with $\varepsilon \in [0,\pi/4)$ are
attracted by $F_0^\varepsilon$ and repeled by $\Fzero$.

In the rest of the space the flow is as follows.
The points above the critical surface $C_{0,1}$ (see
Fig.~\ref{fig5}a)
are driven by the RG transformation to the infinite-temperature
fixed point $F_\infty$. They will reach this point after an infinite
number of RG steps.  In particular, the points at the plane $J=0$ are
mapped into the plane $\varepsilon=0$ in a {\em single}
renormalization
step, where they remain during their migration to $F_\infty$.  (The
points with
$J<0$ go to the previous quadrant $-\pi/2\leq\varepsilon <0$.)
The points below $C_{0,1}$
are swallowed up by the black hole region in a {\it finite} number
of RG steps.
That means that the part of the black-hole region $B_{2k}$ located
below the surface $C_{0,1}$
grows with the order $k$ of the decimation procedure. For
instance, the corresponding black-hole region for the transformation
$T \rightarrow T^3$ (double decimation) contains, besides all the
points belonging
to $B_2$, also the points which are mapped into the black hole
$B_2$ by a single decimation.
(In Section \ref{ablack} we shall see that the part of $B_{2k}$ above
the surface $C_{0,1}$ decreases in size with $k$.)
Finally, the points on the transition
surface $C_{0,1}$ remain there after renormalization.
There are three domains of
attraction: a) if $J = 2J_1$ ($\varepsilon = \pi/2$) the flow goes
towards $F_c$; b) if $J < 2J_1$ ($\varepsilon = \pi/2$), towards
$F_c^\dagger$; and c) if $J > 2J_1$, toward $F_c^\ast$.
In all cases, an infinite number of RG steps are needed to
arrive at the corresponding fixed point.
%

\section{Analytic Extensions of the 4-State Clock Model}
\label{ablack}

As explained in the last section, at every point belonging to the
region $B_2$ one has to extend \reff{decimation} analytically to the
complex
$(J,J_1,\varepsilon)$ space\footnote{ We thank Alexei Morozov for
pointing out the interest of this question.}. However, to
have a real free energy this extension must be performed in such a way
that the transfer matrix remains Hermitean.
This amounts to a suitable definition of the complex logarithm.
Alternatively, we can ask ourselves how many hermitean extensions of a
matrix $T$ of the form \reff{tmatrix} can be constructed.  After some
straightforward algebra we conclude that there are two of them,
which can be labelled by an integer $m$ taking the values 0 and 1.
The transfer matrix now reads
\begin{equation}
T(m)  = \left(
\begin{array}{cccc}
e^{J+J_1}        &   e^{-i\varepsilon - J_1}    &
(-1)^m e^{-J+J_1}       &   e^{i\varepsilon - J_1}     \\
e^{i\varepsilon - J_1}  &    e^{J+J_1}           &
e^{-i\varepsilon - J_1} &   (-1)^m e^{-J+J_1}          \\
(-1)^m e^{-J+J_1}       &   e^{i\varepsilon - J_1}     &
 e^{J+J_1}        &   e^{-i\varepsilon - J_1}    \\
e^{-i\varepsilon - J_1} &   (-1)^m e^{-J+J_1}          &
e^{i\varepsilon - J_1}  &    e^{J+J_1}
\end{array}
\right)
\label{tnm}
\end{equation}
%
%
where $(J,J_1,\varepsilon) \in \mbox{{\bf R}}^3$. This implies that
the couplings which appear in the Hamiltonian \reff{clock} must be
replaced by the following quantities
\begin{equation}
\begin{array}{lll}
 J            & \rightarrow &  J   - {i m \pi \over 2} \\
 J_1          & \rightarrow &  J_1 + {i m \pi \over 4} \\
 \varepsilon  & \rightarrow &  \varepsilon               \\
 A            & \rightarrow &  A   + {i m \pi  \over 4}
\end{array}
\label{complex}
\end{equation}
%
%
where $A$ is the real coupling associated to the identity operator.
Notice that $\varepsilon$ always remains real.
%
%
The case $m = 0$ corresponds to the original clock model \reff{clock},
and
$m = 1$ to its analytic continuation with coupling constants
carrying an imaginary part.  This last model will be called the
{\em extended} clock model and quantities related with it will be
denoted
with a bar. In Ref.~\cite{clock3} the parameter $m$ was not
considered.

In order to study the phase structure of this extended model, one
has to compute the eigenvalues of the new transfer matrix
$\overline{T} = T(1)$ [given in \reff{tnm}]:
\begin{equation}
\begin{array}{l}
  \bar{\lambda}_0 =
2e^{J_1} \sinh J + 2 e^{-J_1}\cos \varepsilon
  \\
  \bar{\lambda}_2 =
2e^{J_1} \sinh J - 2 e^{-J_1}\cos \varepsilon
  \\
  \bar{\lambda}_1 =
2e^{J_1} \cosh J + 2 e^{-J_1}\sin \varepsilon
  \\
  \bar{\lambda}_3 =
2e^{J_1} \cosh J + 2 e^{-J_1}\sin \varepsilon.
\end{array}
\end{equation}
We can relate easily these ones with those given by \reff{eigenvalues}
\begin{equation}
   \bar{\lambda}_m(J,J_1,\varepsilon) =
        \lambda_{m+1}(J,J_1,\varepsilon+\pi/2)\;.
\end{equation}
This relation implies that the phase structure of the
extended model can be obtained from that of the original one ($m = 0$)
by reflecting Fig.~\ref{fig5}a with respect to the plane
$\varepsilon = \pi/4$.
%
%
Thus, the RG flow in the extended model is the same as in
the original one, except for a reflection (Fig.~\ref{fig5}b).
In particular, to each fixed point located at $(J,J_1,\varepsilon)$
in the original model (listed in Section 3), corresponds an analogous
fixed point but located at
$(J,J_1,\pi/2 - \varepsilon$).  For the same reason there exists
another black hole
\begin{equation}
\overline{B}_2= \{ (J,J_1,\varepsilon) : e^{4J_1} < \cos 2\varepsilon,\
\varepsilon < \frac{\pi}{4} \ ; \ m = 1 \}
\end{equation}
where an analytic extension of the corresponding RG equations is
needed. The curves of LYT singularities are now
\begin{equation}
\overline{\rm LYT1}= \left\{ (J,J_1,\varepsilon) :
J=0=\varepsilon\right\} \label{barlyt1}
\end{equation}
and
\begin{equation}
    \overline{\rm LYT2}= \left\{ (J,J_1,\varepsilon)\in \overline
    C_{0,1} :  \cosh 2J\,\cos 2\varepsilon = 1
    \right\}, \label{barlyt2}
\end{equation}
which intersect at
\begin{equation}
\overline P_{\rm LYT} \;=\; (J=0\,, J_1=0\,, \varepsilon=0).
\label{barplyt}
\end{equation}

The new feature is the flow of points
inside the black-hole regions $B_2$ and $\overline B_2$
from one model to the other.  To study it
we need the RG equations in the slightly higher-dimensional space
$(J,J_1,\varepsilon,m)$, for a transfer matrix $T(m)$ of the form
\reff{tnm}. These are
\begin{eqnarray}
J' &= & \frac{1}{2} \log  \left\{ (-1)^{m+m'}
        \frac{1+e^{4J_1}\cosh 2J }
             {e^{4J_1}+ (-1)^m \cos 2\varepsilon}  \right\}
\nonumber \\
J_1'&=& \frac{1}{4} \log  \left\{ (-1)^{m+m'}
        \frac{ (1+ (-1)^m e^{-4J_1}\cos 2\varepsilon)
               (1+e^{4J_1}\cosh 2J) }
             {2 (\cosh 2J + (-1)^m \cos2\varepsilon)}  \right\}
\label{primes} \\
\varepsilon'&=& \frac{1}{2} \arccos  \left\{
        \frac{ (-1)^m + \cosh 2J \cos 2\varepsilon }
             {\cosh 2J+(-1)^m \cos2 \varepsilon}       \right\},
\nonumber
\end{eqnarray}
where the renormalized integer $m'$ must be chosen in such a way
that there exist real numbers $J'$ and $J_1'$ satisfying
\reff{primes}. This choice is unique.

Consider the original clock model ($m = 0$). If $(J,J_1,\varepsilon)$
does not belong to the black hole $B_2$, then it follows from
\reff{primes}
that $m' = 0$. The flow in this region was described in detail in
Section 3. However
if $(J,J_1,\varepsilon) \in B_2$ then one must choose $m' = 1$; so the
renormalized Hamiltonian belongs to the extended model. By the same
arguments, if $m = 1$ the renormalized Hamiltonian will have $m' = 1$
whenever we start at a point not belonging to $\overline{B}_2$, but
$m'=0$ if $(J,J_1,\varepsilon) \in \overline{B}_2$.
Thus, the analytic extension of the extended model is the original one.

Furthermore, the decimation transformation sends the points inside a
black hole to a {\em different side} of the critical surface:  A point
$(J,J_1,\varepsilon,m=0)\in B_2$ placed {\em below} $C_{0,1}$ is
mapped to a
point $(J',J'_1,\varepsilon',m'=1)\not\in\overline B_2$ located
{\em above} $\overline{C}_{0,1}$.  Such a point is eventually
attracted by the
fixed point $\overline{F}_\infty$, except if it belongs to the line
$J = \infty$ in which case is attracted by the line
$\overline{F}_0^\varepsilon$
(see Figs.~\ref{fig6}a and \ref{fig6}b).
On  the other hand, if $(J,J_1,\varepsilon,m=0)$ belongs to the part
of the black hole on top of $C_{0,1}$, it is mapped to a point outside
$\overline B_2$ and {\em below} $\overline C_{0,1}$.  Such a point
ends up, after a finite number of iterations, inside $\overline B_2$ and
below $\overline C_{0,1}$, so it is returned to the $m=0$ model outside
the black hole $B_2$ and above $C_{0,1}$.  The attractor for such a point
is therefore $F_\infty$.

In addition, a point on the critical surface $C_{0,1}$ and inside
$B_2$ goes to a point on the critical surface $\overline C_{0,1}$ of
the
extended model.  This image point is outside $\overline B_2$, so it is
eventually attracted by $\overline F_c^\ast$

In particular, all this implies that when the order $k$ of the
decimation
increases, the part of the black hole $B_{2k}$ on top of $C_{0,1}$
{\em decreases} in size, with its boundary approaching $C_{0,1}$.
Complementarily, the part of $B_{2k}$ located below $C_{0,1}$
increases
in size, with the boundary also approaching $C_{0,1}$ for large $k$.
Therefore, in the limit of decimations of large order $k$, the
black-hole
region occupies the whole region situated below $C_{0,1}$, and the
boundary $\partial B_{2k}$ asymptotically coincides with this
critical
surface.  All these boundaries $\partial B_{2k}$ contain the curve
LYT2 defined in \reff{lyt2}, and they ``turn'' around it when they
approach $C_{0,1}$ as $k\to\infty$.  Moreover, the curve LYT2 is the
intersection of each $\partial B_{2k}$ with $C_{0,1}$ and it divides
$\partial B_{2k}$ in two parts:  a part inside the black hole $B_{2k}$
(above the curve LYT2 in Fig.~\ref{fig5}a), and a part outside it.
In Section \ref{pato} we discuss other important features of the RG
transformation at this curve.

Analogous considerations hold for the extended ($m=1$) model.  In
particular, points in $\overline B_2$ and below $\overline C_{0,1}$
are attracted by the fixed points $F_\infty$ or $F_0^\varepsilon$ of
the original model; while points in $\overline B_2$ and above
$\overline C_{0,1}$ are attracted by $\overline F_\infty$.
The points in $\overline B_2 \cap \overline C_{0,1}$ are attracted by
$F^\ast_c$.

Finally, the points on the transition surface $C_{1,-3}$ are attracted
by the fixed
point $\overline{F}_c^\dagger$, except those belonging to the curve
$C_{1,-3} \cap C_{0,1}$ which flow to $\overline{F}_c$.

This concludes the analysis of the flow at the non-pathological points,
both in the original and the extended model.
We observe no trace of ambiguities: to each Hamiltonian there
corresponds a unique renormalized Hamiltonian given by \reff{primes}.
Besides, all fixed points are reached after an infinite number of RG
steps,
except $\Fdagger$ and $\overline\Fdagger$ which act as very strong
attractors (attract in one RG step) for the points on the lines
$J=0, J_1=\infty, m=0$ and $J=0, J_1=\infty, m=1$ respectively.

What is left is the study of the pathological regions:  the boundaries
$\partial B_2$ and $\partial\overline B_2$---or, more generally,
$\partial B_{2k}$ and $\partial\overline B_{2k}$---of the black-hole
regions, and the line LYT1.  This is the subject of next section.

\section{Pathologies in the 4-State Clock Model}
\label{pato}

Griffiths and Pearce \cite{griffiths} were, to our knowledge, the first
to point out possible ``peculiarities'' for RG
transformations at a first-order phase transition.  This pioneer
call of attention was followed by numerical results and tentative
arguments \cite{blote,lang,tony_z2,hasenfratz,decker} that seemed to
indicate the possibility of discontinuity and multivaluedness of the
flow at the coexistence points.  The theorems proved in \cite{unpub}
have ruled out such lack of smoothness for real interactions and
compact single-spin space, and have left the more subtle phenomenon of
non-Gibbsianness \cite{sokal,unpub,israel} as the main pathology for
real-space renormalization.

For real interactions, the class of Gibbsian measures is exactly
characterized by the class of quasilocal and uniformly non-null
measures \cite{kozlov}.  Roughly speaking, quasilocality means that
the {\em direct} influence of far away spins on a given spin
$\vec{s}_{x_0}$ is small.  That is, if we take a set $\Lambda$ centered
in $x_0$ and {\em fix} the spins $\vec{s}_x \in \Lambda$, $x \neq x_0$,
then the change of the expected value of $\vec{s}_{x_0}$ with the
boundary conditions outside $\Lambda$ is vanishingly small as the
diameter of $\Lambda$ goes to infinity.
On the other hand, a necessary condition for a measure to be uniformly
non-null is that every open set of configurations have nonzero measure.
For example, the measures obtained for spin systems in the
zero-temperature
limit are typically {\em not} uniformly non-null as they are
concentrated on the configurations which minimize the
energy of the system.
For general {\em complex} interactions the theory of Gibbs measures
is not well developed, but in any case the exponential form of the
Boltzman weights implies that uniform non-nullness is also a necessary
condition for Gibbsianness.

The non-Gibbsianness of the renormalized measure exhibited in
\cite{sokal,unpub,israel} for Ising models at or close to a
first-order
phase transition, is a consequence of lack of quasilocality.  In the
present case, however, the origin of the pathologies is rather
related
to the loss of uniform non-nullness.  The pathologies occur when there
are renormalized models for which some matrix elements of the transfer
matrix \reff{tnm} vanishes.
For instance, if the transfer
matrix is diagonal it would imply that the corresponding measure is
concentrated on constant configurations.
In terms of coupling constants,  measures that are not
uniformly-non-null
correspond to manifolds where $J$ or $J_1$ take an infinite value,
and the pathological points are those mapped into such manifolds by
a finite number of iterations of the RG transformation.

We remark that the ``pathological'' character of a RG transformation
at these points only appears if we insist on finding a renormalized
Hamiltonian.  {\em The renormalized transfer matrix is always well
defined}\/.
This remark is equivalent to the fact, emphasized in
\cite{sokal,unpub},
that renormalization transformations (for real interactions) are
always
well-defined as maps between probability measures.  It is only at the
level of Hamiltonians that the induced transformation can become sick.

To see at which points the single-spin decimation is pathological,
we have to look at the matrix elements of
$T^2$ obtained from \reff{tnm}. These are:
\begin{eqnarray}
(T^2)_{q,q}   &=& 2 \left( e^{2J_1} \cosh 2J + e^{-2J_1} \right)
            \nonumber \\
(T^2)_{q,q+1} &=& 2 \left( e^{J_1  - i \varepsilon} +
                    (-1)^m e^{-J_1 + i \varepsilon} \right)
              = (T^2)_{q,q+3}^{*} \\
(T^2)_{q,q+2} &=& 2 \left( (-1)^m e^{2J_1}  +
                                  e^{-2J_1} \cos 2\varepsilon \right).
            \nonumber
\end{eqnarray}
Zeros of any of these elements determine points that are pathological
for a {\em single} RG map.
The diagonal elements are always non-vanishing; however, the
off-diagonal ones can vanish at some points of the coupling
constant space.
On the one hand, $(T^2)_{q,q+2} = 0$ at the boundaries $\partial B_2$
and $\partial\overline{B}_2$ of the black hole
region, and $(T^2)_{q,q+1} = (T^2)_{q,q+3} = 0$ at
the lines LYT1 and $\overline{\rm LYT1}$.
The matrix $T^2$ is diagonal at $P_{\rm LYT}$ and
$\overline P_{\rm LYT}$.

It is interesting to follow the flow of the points at these
pathological
regions.  For concreteness we analyze the original ($m=0$) model; the
discussion for the extended model is analogous.
We see that all the points in $\partial B_2$ are mapped into the line
$J'=\infty, J'_1=-\infty$:  The points above LYT2 go onto the segment
$0<\varepsilon'<\pi/4$ (i.e. outside the black-hole region $B_2$),
while those below LYT2 end up at the complementary segment
$\pi/4<\varepsilon'<\pi/2$ (deep inside $B_2$).  All the points in
LYT2 are mapped to the point with $\varepsilon'=\pi/4$, i.e. to the
non-Gibbsian
fixed point $F^\ast_c$.  Applying a further iteration of the
transformation
one can see (using, for instance, the expresion for $T^4$ obtained via
\reff{tm}), that the points with $\varepsilon'\neq \pi/4$ recover
finitely-valued couplings:  Those with $0<\varepsilon'<\pi/4$ are
mapped
to points above $C_{0,1}$ and flow towards $F_\infty$; while those with
$\pi/4<\varepsilon'<\pi/2$ are mapped
to points above $\overline C_{0,1}$ ($m=1$ model) and flow towards
$\overline F_\infty$.  On the other hand, the points in LYT2 remain in
$F^\ast_c$ (it is a fixed point !) and so they lead always to
non-Gibbsian renormalized measures.
This non-Gibbsianness is
the result of the strong attraction that a non-Gibbsian fixed point
has on the
curve LYT2 (in the sense of attracting it in finitely many---in fact
one---steps).

The points LYT1 are also victims of a strong attraction:  In one
decimation transformation the half-line $J=0,J_1>0,\varepsilon=\pi/2$
is mapped into the plane $J_1=\infty$, while the other
half-line---$J=0,J_1<0,\varepsilon=\pi/2$---is mapped into the plane
$\overline J_1=\infty$.  As these planes are RG-invariant and formed
by non-Gibbsian models, those points remain  non-Gibbsian under
iterations of renormalization group transformations.
It is suggestive to observe that, at the level of coupling constants,
the RG transformations exhibit discontinuities and multivaluedness at
LYT1.  For example, points arbitrarily close to the line LYT1 and
contained in the plane $J=0$ are renormalized into points with
$\varepsilon'=0$, while equally close points but contained in the plane
$\varepsilon=\pi/2$ remain with $\varepsilon'=\pi/2$.  This shows an
explicit discontinuity.  Moreover, the value of $\varepsilon'$ given in
\reff{decimation} is {\em undefined} at the line LYT1:  By approaching
this line (``preparing the system'') from different directions
$J(\varepsilon)$ one can obtain any value of $\varepsilon'$.  This is
an instance of multivaluedness.  In this sense, the ``most''
pathological
point is $P_{\rm LYT} = {\rm LYT1}\cap{\rm LYT2}$.  Its renormalized
transfer matrix is diagonal and, moreover, {\em both} $\varepsilon'$
and
$J_1'$ are undefined for this point, and can be made to take any
value by approaching it in different ways.

Again we emphasize that this discontinuity and multivaluedness
appears because we are trying to follow finitely-valued couplings,
and ``truncating'' the infinitely-valued ones.  There is no
ambiguity
or lack of smoothness at the level of renormalized transfer matrices.
Analogously, non-Gibbsianness prevents us from linearizing the
transformation {\em as a function of the couplings} around the
critical
fixed points $F_c^\ast$ and $F_c^\dagger$, and hence the matrices
$L_{F_c^\ast}$ and $L_{F_c^\dagger}$ are not defined.  The
linearization
is perfectly possible for the transformation as a function of the
entries of the transfer matrix.

Other pathologies of the single-spin decimation show up if we look
for zeroes among the entries of higher powers of $T$.  Such
pathologies
correspond to the loss of Gibbsianness after a finite number of
renormalizations.  They coincide with single-renormalization
pathologies for decimations of higher order, and they occur at the
successive boundaries $\partial B_{2k}$.  Such pathologies have the
same features as those discussed above:  except at the points
LYT2---contained
in all such boundaries---the Gibbsianness is recovered for good
after a further iteration of the transformation ($\partial B_{2k}
\cap \partial B_{2k'}= {\rm LYT2}$ if $k\neq k'$).
On the other hand the above pathological points depend very much on
the
renormalization group prescription and in this sense thery are not
universal.
However their existence in the espace of local Hamiltonians cannot be
avoided by the choice of a suitable choice of a renormalization group
prescription\footnote{In some cases a description of the
renormalization group
transformation in terms of non-local variables might lead to a
continous flow \cite {Gawedski}.}. In fact the analysis in a  general
decimation prescription
can be obtained from the study of the continuous flow defined
by the infinitesimal transformation $T \rightarrow T^t$ ($t \in $
{\bf R}). This is
given (for $m = 0$) by the following differential equations
\begin{eqnarray}
\dot{J} &=&  (\lambda_1\log\lambda_1 + \lambda_3\log \lambda_3)
                  \frac{\cosh J}{4 e^{J_1} } -
                  (\lambda_0\log \lambda_0 + \lambda_2\log \lambda_2)
                  \frac{\sinh J}{4 e^{J_1} } \nonumber \\
\dot{J}_1 &=&
           \frac{(\lambda_0\log \lambda_0 +  \lambda_2\log \lambda_2)}
                {8 e^{J_1} \cosh J} -
           \frac {(\lambda_0\log \lambda_0 - \lambda_2\log \lambda_2)}
                 {8 e^{-J_1} \cos \varepsilon } -
           \dot{\varepsilon} \frac{\tan\varepsilon}{2} -
           \dot{J} \frac{\tanh J}{2}
\label{dots} \\
\dot{\varepsilon} &=& \frac{e^{J_1}}{4} \left\{
  (\lambda_1\log \lambda_1- \lambda_3\log \lambda_3) \cos \varepsilon -
  (\lambda_0\log \lambda_0 - \lambda_2\log\lambda_2) \sin \varepsilon
                                        \right\} \nonumber
\end{eqnarray}
which are valid only in the reflection-positive region.
These equations are well defined for every point in this domain which
contains
some points of $\partial B_2$. The only pathology of the points of
$B_2$ is
that they are attracted by the line
$J=\infty,J_1=-\infty,\varepsilon\in$
$(\pi/4,\pi/2$] in a finite  ``time'' $t<2$. This feature explains the
especial
behavior of the points of $B_2$ under the action of
$T^2$-renormalization group transformations: they are mapped into the
$m=1$ domain. The singularity arises also in this infinitesimal scheme
when we
integrate \reff{dots} up to a {\it finite} renormalization scale $t$.
For each
$t$ there is a black hole region $B_t$ that changes  as we increase
the scale
$t$.  In fact, as we have seen, every point below the transition
surface belongs
to a region $B_t$ for $t$ large enough because they are attacted  by
the line
$J=\infty,J_1=-\infty,\varepsilon\in $ $(\pi/4,\pi/2$). This implies
that the
flow defined by the differential equations \reff{dots} does not
generate a one-parameter group of global  transformations.

\section{Conclusions}

We believe that the model studied here provides an instructive
illustration
of the behavior of real-space renormalization transformations in the
vicinity of coexistence manifolds.  Admittedly, some of the features
observed could be specifically due to the use in particular of
decimation
procedures, but we expect others to be indications of phenomena of
more general validity.

The present example explicitly exhibits a picture that takes
considerable
more work to verify for real interactions \cite{sokal,unpub,israel}:
the reconciliation of the standard scenario and the presence of
pathologies,
and non-Gibbsianness as a source of an observed discontinuity and
multivaluedness
in the flow of the coupling constants.  In addition, we consider
singularly
suggestive the observations that the worst type of pathologies takes
place at points where all the eigenvalues of the transfer matrix are
doubly degenerate, and where there are singularities of the type
linked, in some studies, to metastability effects.

The analysis performed here can be extended to $q$-state clock models
of higher $q$.  We do not expect any new phenomenon except, perhaps,
in the limit $q\to\infty$ (chain of plane rotors with complex
interactions).
But of course, the challenge is to find a model with {\em real}
interactions
exhibiting, with comparable explicitness, non-Gibbsianness of
renormalized measures and its consequences for computational
schemes (an aspect left unresolved in previous work
\cite{sokal,israel,griffiths}).  Such an example, however, may not be
easy to construct given the rather subtle character of
non-quasilocality.

\section*{Acknowledgements}

We thank Antonio Gonz\'alez-Arroyo and Alexei Morozov
for useful discussions, Aernout C. D. van Enter for carefully reading
and critizicing the manuscript, and
Prof. Michael Fisher for correspondence. We also thank to Prof. Joel
Lebowitz for bringing Ref.~\cite{Gallavotti-Lebowitz} to our
attention.  R.~Fern\'andez and J.~Salas thank the members of
Departamento de F\'{\i}sica
Te\'orica of the Universidad de Zaragoza and of the Universidad
Aut\'onoma de Madrid (R.F.) for their warm hospitality
during the completion of this work. We also acknowledge CICyT for
partial finantial support  (grants AEN90-0029 and AEN90-0030).

\newpage

\section*{Figure Captions}

\noindent  {\bf Figure 1:}
Eigenvalues $\lambda_k$ ($k = 0, \ldots,3$) of the
transfer matrix (5) as a functions of $\varepsilon$ for $J = 1.5$ and
$J_1 = 0.5$.
\medskip

\noindent  {\bf Figure 2:}
Transition surfaces
and LYT points
for the model in the region
$J\geq 0, 0\leq\varepsilon\leq\pi/2$.  The intersection with the
reflection-positivity interface of Fig.~\protect\ref{fig3} is also
shown.
\medskip

\noindent  {\bf Figure 3:}
Reflection positivity interface \protect\reff{rp} for the
model. We have also drawn some of the fixed points, the
renormalized trajectory joining $F_c$ and $F_c^\ast$ and the black
hole region \protect\reff{black}.
\medskip

\noindent  {\bf Figure 4:}
Transition surface outside the black hole region
\protect\reff{black}. The contour of the reflection positivity
interface is also drawn for clarity.
\medskip

\noindent  {\bf Figure 5:}
Renormalization group flow in the reflection positive region:
(a) for the original model, and (b) for the extended one. Fixed points
and some of the renormalized trajectories are also despicted. The
black
hole region is despicted in the right side of (a) and in the left one
of (b).
\medskip

\noindent  {\bf Figure 6:}
Universality classes (a) for $\varepsilon \in (0,\pi/2)$,
and (b) $\varepsilon = \pi/2$. The fixed point which attracts all
points belonging to a region is written inside it. Notice that
the surface $C_{0,-3}$, defined by the condition
$\lambda_0=-\lambda_3 > \max (|\lambda_1|,|\lambda_2|)$, is a critical
one in the region $J<0$ (See Eq. (10)). The dashed line represents the
boundary $\partial B_2$ and the dash-dotted line the LYT1 points.

%
%
\newpage
\begin{figure}
\mbox{}
\caption{}
\label{fig1}
\end{figure}

\begin{figure}
\mbox{}
\caption{}
\label{fig2}
\end{figure}

\begin{figure}
\mbox{}
\caption{}
\label{fig3}
\end{figure}

\begin{figure}
\mbox{}
\caption{}
\label{fig4}
\end{figure}

\begin{figure}
\caption{}
\mbox{}
\label{fig5}
\end{figure}

\begin{figure}
\mbox{}
\caption{}
\label{fig6}
\end{figure}


\newpage
\begin{thebibliography}{99}

\bibitem{wilson}
K. G. Wilson, {\it Rev. Mod. Phys.} {\bf 47} (1975) 773.

\bibitem{wilson-kogut}
K. G. Wilson and J. Kogut, {\it Phys. Repts.} {\bf 12} (1974) 75.

\bibitem{nienhuis}
B. Nienhuis and M. Nauenberg, {\it Phys. Rev. Lett.} {\bf 35}, (1975)
477.

\bibitem{klein}
W. Klein, D. J. Wallace and R. K. P. Zia, {\it Phys. Rev. Lett.} {\bf
37}, (1976) 639.

\bibitem{fisher}
M. E. Fisher and A. N. Berker, {\it Phys. Rev.} {\bf B26}, (1982) 2507.

\bibitem{blote}
H. J. Bl\"ote and R. H. Swendsen, {\it Phys. Rev. Lett.} {\bf 43}
(1979) 799.

\bibitem{lang}
C. B. Lang, {\it Nucl.  Phys.}  {\bf B280 } [FS18] (1987) 255.

\bibitem{tony_z2}
A. Gonz\'alez-Arroyo, M. Okawa and Y. Shimizu, {\it Phys. Rev. Lett.}
{\bf 60} (1988) 487.

\bibitem{hasenfratz}
A. Hasenfratz and P. Hasenfratz, {\it Nucl. Phys.}  {\bf B295 }
[FS21] (1988) 1.

\bibitem{decker}
K. Decker, A. Hasenfratz and P. Hasenfratz, {\it Nucl. Phys.}
{\bf B295 } [FS21] (1988) 21.

\bibitem{sokal}
A. C. D. van Enter, R. Fern\'andez and A. Sokal, {\it Phys. Rev. Lett.}
{\bf 66} (1991) 3253.

\bibitem{unpub}
A. C. D. van Enter, R. Fern\'andez and A. Sokal, in preparation.

\bibitem{israel}
R. B. Israel, in {\it Random Fields} Vol. II, Eds. J. Fritz,
J. L. Lebowitz, D. Sz\'asz. North Holland, Amsterdam (1981).

\bibitem{griffiths}
R. B. Griffiths and P. A. Pearce, {\it Phys. Rev. Lett.} {\bf 41} (1978)
917; {\it J. Stat. Phys.} {\bf 20} (1979) 449.

\bibitem{tony-salas}
A. Gonz\'alez-Arroyo and J. Salas, {\it Phys. Lett.} {\bf B261} (1991)
415.

\bibitem{wilson2}
K. G. Wilson, in {\it Recent Developments in Gauge
Theories}, eds. G.'t Hooft {\it et al}. Plenum, New York (1980).

\bibitem{asorey}
M. Asorey and J. G. Esteve, {\it J. Stat. Phys.} {\bf 65} (1991) 483.

\bibitem{affleck}
I. Affleck, {\it Nucl. Phys.} {\bf B257} [FS14] (1985) 397.

\bibitem{haldane}
F. D. Haldane, {\it Phys. Rev. Lett.} {\bf 50} (1983) 1153.

\bibitem{affleck91}
I. Affleck, {\it Phys. Rev.} {\bf B43} (1991) 3215.

\bibitem{pruisken}
A. M. M. Pruisken, {\it Phys. Rev.} {\bf B15} (1985) 2636.

\bibitem{levine}
H. Levine and S. B. Libby, {\it Phy. Lett.} {\bf B150} (1985) 182.

\bibitem{chiralpotts1}
H. Au-Yang, B. M. McCoy, J. H. H. Perk, S. Tang and M. L. Young,
{\it Phys. Lett.} {\bf A123} (1987) 219.

\bibitem{chiralpotts2}
R. J. Baxter, J. H. H. Perk and H. Au-Yang, {\it Phys. Lett.} {\bf A128}
(1988) 138.

\bibitem{ruelle}
D. Ruelle, {\it Statistical Mechanics}\/, Benjamin (1969);
{\it Thermodynamic Formalism}\/, Addison Wesley (1978).

\bibitem{israelconvexity}
R. B. Israel, {\it Convexity in the Theory of Lattice Gases},
Princeton University Press (1979).

\bibitem{Lee-Yang}
T. D. Lee and C. N. Yang, {\it Phys. Rev.} {\bf 87} (1952)
404; 410.

\bibitem{Griffiths}
P. J. Kortman and R. B. Griffiths, {\it Phys. Rev. Lett.} {\bf 27}
(1971) 1439.

\bibitem{Fisher}
M. E. Fisher, {\it Phys. Rev. Lett.} {\bf 40} (1978) 1610;
{\it Prog. Theor. Phys. Suppl.} {\bf 69} (1980) 14.

\bibitem{Cardy}
J. L. Cardy, {\it Phys. Rev. Lett.} {\bf 54} (1985) 1354.

\bibitem{Gallavotti-Lebowitz}
G. Gallavotti, J. Lebowitz, {\it Physica} {\bf 70} (1973) 219.

\bibitem{isakov}
S. N. Isakov, {\it Comm. Math. Phys.} {\bf 95} (1984) 443.

\bibitem{milos}
M. Zahradnik, {\it J. Stat. Phys.} {\bf 47} (1987) 725.

\bibitem{galsol}
G. Gallavotti and S. Miracle-Sol\'e, {\it Comm. Math. Phys.} {\bf 7}
(1968) 274.

\bibitem{isr}
R. B. Israel, {\it Comm. Math. Phys.} {\bf 50}
(1976) 245.

\bibitem{dobmar}
R. L. Dobrushin and M. R. Martirosyan, {\it Teor. Mat. Fiz.} {\bf 74}
(1988) 16.

\bibitem{teoremas}
H. O. Georgii, {\it Gibbs Measures and Phase Transitions}, de Gruyter,
Berlin-New York (1988) and references therein.

\bibitem{asorey-esteve}
M. Asorey, J.G. Esteve and A. F. Pacheco, {\it Phys. Rev} {\bf D27}
(1983) 1852.

\bibitem{borgs}
C. Borgs and J. Z. Imbrie, {\it Comm. Math. Phys.} {\bf 123}
(1989) 305.

\bibitem{Fisher-Nelson}
M. E.  Fisher, D. R.  Nelson, {\it Ann. Phys. (N.Y.)} {\bf 91} (1975)
226.

\bibitem{clock3}
M. Asorey, J. G. Esteve and J. Salas, DFTUZ 91.32 preprint.

\bibitem{kozlov}
O. K. Kozlov, {\it Probl. Inform. Transmission} {\bf 10} (1974) 258.

\bibitem{minlossinai}
R. A. Minlos and Ya. G. Sinai, {\it Theor. Mat. Fiz.} {\bf 2} (1970) 230;
translation {\it Theor. Math. Phys.} {\bf 2} (1971) 167.

\bibitem{Gawedski}
K. Gawedzki, R. Kotecky and A. Kupiainen, {\it Comm. Math. Phys.} {\bf 47}
(1987) 701.

\end{thebibliography}
\end{document}